\documentstyle[prl,aps,epsfig]{revtex}

\begin{document}
\title{Evidence of
exactness of the mean field theory in the 
nonextensive regime of
long-range spin models} 
\author{S.A. Cannas$^{1*}$, A.C.N. de
Magalh\~aes$^{2}$ and 
F.A. Tamarit$^{1*}$}

\address{
$^{1}$Facultad de
Matem\'atica, Astronom\'\i a y F\'\i sica,
Universidad Nacional de C\'ordoba, Ciudad Universitaria, 5000
C\'ordoba, Argentina \\
$^{2}$Centro Brasileiro de Pesquisas F\'\i sicas, Xavier Sigaud 150, 22290-180 Rio de Janeiro-RJ, Brazil}

\date{\today}
\maketitle

\begin{abstract}
The $q$-state Potts model
with long-range interactions that decay as
$1/r^\alpha$ subjected to an
uniform magnetic field on $d$--dimensional
lattices is analized 
for
different values of $q$ in the nonextensive
regime $0\leq\alpha\leq d$. We
also consider the two dimensional
antiferromagnetic Ising model with the
same type of interactions. The mean
field solution and Monte Carlo
calculations for the equations of state for
these models are compared. We
show that, using a derived scaling 
which properly describes the
nonextensive thermodynamic
behaviour, both types of calculations show an
excellent agreement in all the
cases here considered, except for $ \alpha=d
$.
These results allow us to extend to nonextensive magnetic models
with
$0\leq\alpha<d$
a previous conjecture which states that the mean field
theory is exact 
for the Ising
one.

\end{abstract}


\section{Introduction}

Microscopic pair interactions that decay slowly with the distance
$r$
between particles appear in different physical systems. Typical
examples are
gravitational and Coulomb interactions, where the potential
decays as $1/r$.
Several other important examples can be found in condensed
matter, such as
dipolar (both electric and magnetic) and
Ruderman-Kittel-Kasuya-Yosida
(RKKY) interactions, both proportional to
$1/r^3$. Effective interactions
with a power law decay $1/r^\alpha$, with
some exponent $\alpha\geq 0$,
appear also in other related problems such as
critical phenomena in highly
ionic systems\cite{Pitzer}, Casimir forces
between inert uncharged particles
immersed in a fluid near the critical
point\cite{Burkhardt} and phase
segregation in model
alloys\cite{Giacomini}.

It is known that some of these systems can exhibit
nonextensive
thermodynamic behaviour (see Refs.\cite{Goldenfeld,Cannas1}
and references
therein). In other words, for small enough values of the
ratio $\alpha/d$
the free energy $F = - \ln{Z}/\beta$, with $Z\equiv
Tr\,
exp(-\beta\,H)$ ($H$ being the Hamiltonian of the system, $d$
the
dimensionality and $\beta\equiv 1/ k_BT$), grows faster than the number
$N$ of
microscopic elements when $N\rightarrow \infty$, and the
thermodynamic limit
is not well defined.

In a recent
communication\cite{Cannas1} two of us analized the
thermodynamics
associated with the long range (LR) ferromagnetic Ising
Hamiltonian

\begin{equation}
H = -\sum_{(i,j)} J(r_{ij})\; S_i S_j
\qquad
\text{($S_i=\pm 1$\,\, $\forall$
$i$)},
\label{H1}
\end{equation}

with

\begin{equation}
J(r_{ij})=\frac{J}{r_{i
j}^\alpha}
\qquad \text{($J>0$; $\alpha \geq
0$)}
\label{Jrij}
\end{equation}

\noindent where $r_{ij}$ is the distance
(in crystal units) between
sites $i$ and $j$, and where the sum
$\sum_{(i,j)}$ runs over all
distinct pairs of sites on a $d$-dimensional
hypercubic
lattice. It was  shown\cite{Cannas1} that the quantities per
particle:
free energy $f$, 
internal energy $u$, entropy $s$ and
magnetization $m$ of a finite 
model of $N$ spins behave according
to
Tsallis conjecture\cite{Tsallis} for $N \gg 1$. These quantities 
present, in the presence of
an external magnetic field $h$,  the following
asymptotic scaling behaviours:

\begin{eqnarray}
u(N,T,h) &\sim&   N^*\,
u'(T/N^*,h/N^*)  \label{U} \\
f(N,T,h) &\sim&   N^*\, f'(T/N^*,h/N^*)
\\
s(N,T,h) &\sim&   s'(T/N^*,h/N^*)  \\
m(N,T,h) &\sim&   m'(T/N^*,h/N^*)
\label{M}
\end{eqnarray}

\noindent for all $\alpha\geq 0$, where the
functions $u'$, $f'$, $s'$ and
$m'$ are the corresponding quantities
associated with the same model but
with rescaled coupling
$J'(r_{ij})=J(r_{ij})/N^*$ (these functions 
are independent of the system
size $N$) and the function $N^*(\alpha)$
is defined
as

\begin{equation}
N^*(\alpha) = \frac{1}{1-\alpha/d}
(N^{1-\alpha/d}-1)
\label{N*}
\end{equation}

\noindent which behaves, for
$N\to \infty$, as 

\begin{equation}
N^*(\alpha) \sim
\left\{
\begin{array}{ll}
\frac{1}{\alpha/d-1}& \;\;\;\text{for $\alpha/d >
1$} \nonumber\\
\ln{N} & \;\;\;\text{for $\alpha/d =
1$}\\
\frac{1}{1-\alpha/d} N^{1-\alpha/d}& \;\;\;
\text{for $0\leq\alpha/d<
1$}
\end{array} \right.
\label{n*}
\end{equation}

\noindent For $\alpha>d$
the thermodynamic
functions per site do not depend on $N$ and the system is
extensive ({\it i.e.},
the thermodynamic limit exists). When
$0\leq\alpha\leq d$, $N^*(\alpha)$
diverges for  $N\rightarrow\infty$ and
the system is non-extensive. It was also
presented numerical
evidence\cite{Cannas1} that for $d=1$ the mean field
(MF) theory becomes
{\bf exact} when $0\leq\alpha<d$.
This led two of us
  to conjecture that
the mean field theory might be exact for the
nonextensive Ising model.

In
this work we extend the previous analysis and present new evidences of
the exactness of the MF theory for $0\leq\alpha<d$
 for the one dimensional ferromagnetic $q$-state LR Potts
model for different values of $q$ (including the first order phase
transition predicted by MF theory for $q>2$) and also for the  two-
dimensional antiferromagnetic LR Ising model in an external field.We, 
thus, conjecture that the MF theory might be
exact for any nonextensive magnetic model excluding the borderline
case $\alpha=d$, where  there are probably corrections
to the MF results (some previous experimental results \cite{Nielsen}
on dipolar ferromagnetic materials do not exclude this possibility).

The outline of this paper is the following. In section II we analize the
ferromagnetic q--state LR Potts model subjected to an
uniform magnetic field $h$. First, we show that the previous analysis
for the Ising model ($q=2$) in the non-extensive region is straightforwardly
extended to the generic $q > 2$ case. Then, in subsection II-A we derive
the mean field solution of this model for arbitrary values of $q$, 
$\alpha$ and $h$, in particular the MF predictions for the LR bond 
percolation which corresponds to the $q\to 1$ and $h\to 0^{+}$ limit. In subsection II-B we compare the MF solution with
our Monte Carlo simulation of the one-dimensional model for $h=0$, $q=2,3$
and $5$ and different values of $\alpha$.
In section III we calculate the mean field solution of the two-dimensional
Ising model with competing LR antiferromagnetic  and short-range
ferromagnetic interactions in an external field and compare them with the
Monte Carlo results of Sampaio et al\cite{Sampaio}. Some comments and
conclusions are presented in section IV.


\section{Potts model with long-range ferromagnetic interactions}

 In this
section we address the LR ferromagnetic $q$ state Potts
model, {\it i.e.},
we consider the Hamiltonian:

\begin{equation}
H = - \frac{1}{2} \;
\sum_{i,j} J(r_{ij})\;
          \delta(\sigma_i,\sigma_j )- h \sum_i
\delta(\sigma_i,1)
    \qquad \text{($\sigma_i=1,2,\ldots,q$, $\forall$
$i$)}
\label{H2}
\end{equation}

\noindent where to each site, $i$, we
associate a spin variable $\sigma_i$,
which can assume $q$ integer values;
the sum $\sum_{i,j}$
runs over all distinct pairs of sites of a
$d$-dimensional lattice  of
$N$ sites ($i\neq j$); $\delta$ is the
Kronecker delta function, $J(r_{ij})$
is given by Eq.(\ref{Jrij}) and $h$
is an external magnetic field in the
$\sigma=1$ direction. The $\alpha
\rightarrow
\infty$ limit corresponds to the first-neighbor model. For
$q=2$ the
$\alpha=0$ limit corresponds, after a
rescaling $J\rightarrow
J/N$, to the Curie-Weiss model.

This model, in its plain formulation
($\alpha\rightarrow\infty$
of Eq.(\ref{H2}))
or in a more general one with
many-body interactions, is at the
heart of a complex network of relations
between geometrical
and/or thermal statistical models, like for example
various
types of percolation, vertex models, generalized resistor and
diode
network problems, classical spin models, etc (see \cite{Tsallis2}
and
references therein).

On the other hand, the Potts model with LR
interactions has been much less studied. In the extensive regime
$\alpha>d$ it
presents a very rich thermodynamic behaviour, even in the
one-dimensional
case\cite{Cannas2,Glumac,Glumac2}. To the best of our
knowledge, no study
has been carried out for the nonextensive regime
$0\leq\alpha\leq d$.

Let us introduce the sums $\phi_i(\alpha)=\sum_{j\neq
i} 1/r_{ij}^\alpha$.
A sufficient condition (and believed to be
necessary\cite{Thompson,Aizenman})
for the existence of the
thermodynamic limit of this system is that

\begin{equation}
\phi(\alpha)=
\lim_{N\rightarrow\infty} \frac{1}{N} \sum_i
\phi_i(\alpha) <
\infty.
\label{phi}
\end{equation}

Let us now consider a d-dimensional
hypercube of side $L+1$ and
$N=(L+1)^d$, and let $i=0$ be the central site
of the hypercube. We have that

\begin{equation}
\phi(\alpha)=\lim_{N\rightarrow\infty}
\phi_0(\alpha).
\label{phi0}
\end{equation}

When $L \gg 1$ ($N \gg 1$) $\phi_{0}(\alpha)$ shows the following
asymptotic behaviour\cite{Cannas1}:

\begin{equation}
\phi_0(\alpha) \sim
C_d(\alpha)\, 2^\alpha\,
N^*(\alpha)
\label{phi02}
\end{equation}

\noindent where $N^*(\alpha)$ is
given by Eq.(\ref{N*}) and $C_d(\alpha)$ is
a continuous
function\cite{Cannas1} of $\alpha$ independent of $N$, with
$C_d(0)=1$ $\forall d$. It can be proved that\cite{Cannas1}:

\begin{equation}
C_1(\alpha) = \left\{
\begin{array}{ll}
         1&   \text{for $0\leq\alpha\leq 1$} \\

\frac{\alpha-1}{2^{\alpha-1}}\zeta(\alpha)&
         \text{for $\alpha>1$}

\end{array} \right. .
\end{equation}

\noindent where $\zeta(x)$ is the Riemann Zeta function. 
From Eqs.(\ref{N*})--(\ref{phi02}) we see that the thermodynamic
limit is well defined for $\alpha>d$ (where the system presents
extensive behaviour), while for $\alpha \leq d$ the system
becomes non-extensive. Following the same procedure as in
Ref.\cite{Cannas1} it can be shown that the scaling behaviours
(\ref{U})-(\ref{M}) of the thermodynamic functions hold $\forall q\geq 2$
and $\forall \alpha\geq 0$. For $q=2$ the system undergoes a second
order phase transition at finite temperature for all $\alpha>d$
when $d\geq 2$\cite{Hiley} and for $1<\alpha\leq 2$
when $d=1$\cite{Dyson}. For $\alpha \rightarrow d^+$, the critical temperature shows the following asymptotic behaviour \cite{Hiley}:

\begin{equation}
k_BT_c \sim J\, \phi(\alpha)
\end{equation}

\noindent For $d\geq 2$  and $\alpha\gg d$ (short-range case\cite{Tsallis2})
there exists a critical value $q_c$ such that the phase transition is a
second order one when $q\leq q_c(d)$ ($q_c=4$ for $d=2$) and a first order one for $q>q_c(d)$. For $d=1$  and $1<\alpha\leq 2$ Monte Carlo simulations\cite{Glumac2} show that, for $q>2$, there is a q-dependent 
threshold value $\alpha_{c}(q)$ such that the transition is of first order
when $\alpha<\alpha_{c}(q)$ and of second order above it.

\subsection{Mean field theory}

In order to develop a mean field version of Hamiltonian
(\ref{H2}) we use
Mittag and Stephen\cite{Mittag} spin representation for
the Potts model,
{\it i.e.}, we associate to each site $j$ a spin variable
$\lambda_j$ which
can take the values $\lambda_j= 1, \omega, \omega^2,
\ldots, \omega^{q-1}$,
where $\omega=e^{2\pi\, i / q}$ is a $q$th root of
unity. In other words, if
the site $j$ is in the state $\sigma$ then
$\lambda_j = \omega^{\sigma-1}$.
Then, using the
property

\begin{equation}
q^{-1}\, \sum_{k=1}^q\, \lambda^k\,
\lambda'^{q-k}
=
\delta(\lambda,\lambda')
\label{suma}
\end{equation}

\noindent we can
rewrite the Hamiltonian (\ref{H2}) as:

\begin{equation}
H = - \frac{1}{2q}
\; \sum_{i,j} J(r_{ij})\; \sum_{l=1}^{q-1}
          \lambda_i^l\,
\lambda_j^{q-l} - \frac{h}{q}  \sum_i
\sum_{l=1}^{q-1} \lambda_i^l \; - \;
C(J,h) \;
\label{H3}
\end{equation}

\noindent where the constant term
$C(J,h)$ is

\begin{equation}
C(J,h) = \frac{1}{2q} \sum_{i,j} J(r_{ij}) \;
+ \; \frac{h N}{q} \; .
\label{constante}
\end{equation}  The fraction of
sites
in the state $\sigma$, $n_\sigma=(1/N)\, \sum_i
<\delta(\sigma_i-\sigma)>$, in
this representation is given
by:

\begin{equation}
n_\sigma=\frac{1}{q}\, \left[ 1+ \sum_{l=1}^{q-1}
\omega^{q-l(\sigma-1)}
\left< \lambda^l \right>
\right],
\label{nsigma}
\end{equation}

\noindent and the order parameter
for a symmetry breaking in the $\sigma=1$
direction is

\begin{equation}
m
= \frac{q\, n_1 -1}{q-1} \, ,
\label{mag}
\end{equation}

\noindent which
can be written, using Eq.(\ref{nsigma}), as:

\begin{equation}
m =
\frac{1}{q-1} \, \sum_{l=1}^{q-1} \left< \lambda^l
\right>.
\label{m1}
\end{equation}

The mean field solution for this model
can be easily found from
the Variational Method in Statistical
Mechanics\cite{Huang} by
using a non--interacting trial Hamiltonian $H_{0}$
given by

\begin{equation}
H_{0} = -\eta \sum_{i=1}^{N} \sum_{l=1}^{q-1}
\lambda_{i}^{l}
\end{equation}

\noindent where $\eta$ is the variational
parameter to be found as a
function of temperature. The variational free
energy $\overline{F}$
is given by 

\begin{eqnarray}
\overline{F} &= &F_{0}
\; + \; \langle H - H_{0} \rangle_{0} \\
   & = & F_{0} \; + \; (\eta
-\frac{h}{q}) \sum_{i} \sum_{l=1}^{q-1}
\langle \lambda_{i}^{l} \rangle_{0}-
\frac{1}{2q} \sum_{i,j} J(r_{ij}) \sum_{l=1}^{q-1}
\langle
\lambda_{i}^{l} \lambda_{j}^{q-l} \rangle_{0} -
C(J,h)
\label{fbarra}
\end{eqnarray}

\noindent where the free energy $F_{0}$ associated with $H_{0}$ is

\begin{equation}
F_{0} =
-\frac{N}{\beta} \ln{ [ \exp{( \beta \eta (q-1))} + 
(q-1)
\exp{(-\beta\eta)}
] }
\end{equation}

\noindent and $\langle \ldots \rangle_{0}$ denotes the canonical average 
using the Boltzmann measure proportional to $\exp{ (-\beta H_{0}) }$.

Using equality (\ref{suma}) one gets that

\begin{equation}
\langle \lambda_{i} \rangle_{0} = \langle
\lambda_{i}^{2} \rangle_{0} =
\dots = \langle \lambda_{i}^{q-1} \rangle_{0}
= m_{0} \qquad
\qquad (\forall i)
\end{equation}

\noindent where the variational order parameter $m_{0}$ (defined 
by an equation similar to Eq.(\ref{m1})) is related to $\eta$ 
through:

\begin{equation}
m_{0} = \frac{
\exp{(\beta \eta q)} -1 }{\exp{(\beta \eta q)} +
(q-1)}
\label{eme0}
\end{equation}

\noindent and from the property $\lambda^{q-l}=(\lambda^{l} )^{*} $ it
follows that

\begin{equation}
\langle \lambda_{i}^{l} \lambda_{j}^{q-l}\rangle_{0} =
m_{0}^{2} \; .
\end{equation}

The minimization condition leads to

\begin{equation}
\eta = \frac{1}{q} [ J \phi(\alpha) m_{0} + h
]
\label{eta}
\end{equation}

\noindent which, combined with Eq.(\ref{eme0}) gives the following
mean field equation for the order parameter $m_{0}$:

\begin{equation}
m_{0}= \frac{\exp{\left[\beta \left(
J\, \phi(\alpha)\, m_{0} +  h\right)
 \right]}-1 }{\exp{\beta\left[ J\,
\phi(\alpha)\, m_{0} +  h \right]}+(q-1)}
\;
.
\label{mfequation}
\end{equation} 

In the  $\alpha\rightarrow\infty$
limit (short range interactions) we have
$\phi(\alpha)\rightarrow z$, $z$
being the coordination number of the
lattice, and we recover,for $h=0$,
Mittag and Stephen's \cite{Mittag} result. For $q=2$
the Hamiltonian
(\ref{H2}) is equivalent to the Ising one with long range
interactions,
providing that $J^{Potts}=2\, J^{Ising}$ and  $h^{Potts}=2\,
h^{Ising}$. In
this case Eq.(\ref{mfequation}) reduces
to

\begin{equation}
\label{magq2}
m_{0}=\tanh{\left[ \beta/2 (J\,
\phi(\alpha) m_{0} + h)\right]}
\end{equation}

\noindent and we recover
the result from Ref.\cite{Cannas1}, {\it i.e.}, the
long range version of
the Curie-Weiss equation which describes a second
order phase transition
for $h=0$ at $k_{B}T_{c}/J^{Ising}\phi(\alpha)=1$.

The variational free energy $\overline{F}$ calculated at the minimum
(condition
(\ref{eta})) gives the following mean field  free energy per site $f$:

\begin{eqnarray}
-\beta f & = & \ln{q} + \frac{\beta J
\phi(\alpha)}{2q} \; + \;
\frac{1}{q} \left\{ \beta h + \frac{1}{2} (q-1)
\beta J \phi(\alpha) 
m_{0}^{2} \right.
\nonumber \\
& & \left. + (q-1)
\beta h m_{0} - [1+(q-1) m_{0}] \ln{[1+(q-1) m_{0}]}
-(q-1) (1- m_{0}) \ln{
(1-m_{0})} \right\}
\label{betaf}
\end{eqnarray}

\noindent where we have used Eqs. (\ref{fbarra})--(\ref{mfequation}).

For $q\geq 3$ and $h=0$ the
transition is of first order and it is easy to verify from Eq.(\ref{betaf}) that at the
critical temperature the order parameter $m$ jumps from zero to the value
$(q-2)/(q-1)$. The critical temperature for $ q\geq3$ is given
by

\begin{equation}
\label{Tc}
k\, T_c/J = \left( \frac{q-2}{q-1} \right)
\frac{\phi(\alpha)}{2\,
\ln{(q-1)}}.
\label{TsobreJ}
\end{equation}

\noindent which recovers, in the $\alpha
\to \infty$ limit, Mittag and Stephen's result\cite{Mittag}.

In the $q\to
1 $ limit, which corresponds to a bond percolation where
the bond
probability occupancy between any two sites $i$ and $j$
is
$p/r_{ij}^{\alpha}$, the order parameter {\em probability
percolation}
$P_{\infty}$ (defined as the probability that a randomly
chosen bond
of an infinite lattice belongs to a cluster of infinite size)
can
be derived from\cite{Kasteleyn}

\begin{equation}
P_{\infty}(p) = 1 \;
+ \; \lim_{\overline{h}\to 0^{+}}
\frac{\partial{}}{\partial{\overline{h}}}
\left\{ \lim_{q\to 1} \frac{\partial{}}{\partial{q}} \lim_{N\to \infty}
\frac{1}{N} \ln{(Z)}] \right\}
\label{Pinfinito}
\end{equation}

\noindent
where $\overline{h} = \beta h$, $Z$ is the partition
function of the Potts
model with coupling constants $J/r_{ij}^{\alpha}$
and $p$ is the first
neighbor bond probability given by $p=1-\exp{(-J/k_B T)}$.

One can easily
show, from Eqs. (\ref{mfequation}), (\ref{betaf})
and (\ref{Pinfinito}),
that the probability percolation $P_{\infty}(p)$
is, as
expected\cite{Aizenman}, exactly the $q\to 1$ and $\overline{h} 
\to 0^{+}$
limit of the order parameter $m_{0}$, namely, 

\begin{equation}
P_{\infty}
= m_{0} (q\to 1, \overline{h} \to 0^{+}) = 
1- exp{ [
-\frac{J\phi(\alpha)}{k_{B}T} m_{0} (q\to 1, \overline{h} 
\to 0^{+})
]}
\end{equation}

\noindent or, in terms of
$p$,

\begin{equation}
P_{\infty}(p;\alpha) = 1 - (1-p)^{\phi(\alpha)
P_{\infty}(p;\alpha)}
\label{Pinfinitodep}
\end{equation}

\noindent from
which it follows that, due to the divergence of 
$\phi(\alpha)$ in the
non--extensive regime, $P_{\infty}(0 < p \le 1;
0 \le \alpha \le d) = 1$
and, hence, the critical probability 
$p_{c}(0 \le \alpha \le d) = 0$ in
agreement with the exact 
result\cite{Aizenman}. 

One can easily prove, from Eq.(\ref{Pinfinitodep}), that the 
percolation probability $P_{\infty}$ of a finite system with
$N$ bonds presents an asymptotic scaling behavior similar 
to Eq.(\ref{M}), namely

\begin{equation}
P_{\infty}(N,p) \sim P_{\infty}'(p^{*}) \qquad
\qquad (N>>1)
\end{equation}

\noindent where $p^{*}$ is the variable $p$
calculated at $T^{*}=T/N^{*}$, namely, 

\begin{equation}
p^{*} = 1 -
\exp{\left(-\frac{JN^*}{k_B T} \right) }
\end{equation}

\noindent and
$P_{\infty}'$ is the probability percolation associated
with the long range
bond percolation whose bond probability occupancies
are $p^{*}/r_{ij}$.
Using this rescaled variable $p^*$, the MF order
parameter equation
becomes, for $N>>1$ (see Eq.(\ref{phi02})):

\begin{equation}
P_{\infty}(p^{*};\alpha) = 1 - (1-p^{*})^{C_{d}(\alpha) 2^{\alpha}
P_{\infty}(p^{*};\alpha)}
\end{equation}

\noindent which leads, for
different values of $\alpha \in [0,d]$,
to monotonously increasing distinct
order parameters as $p^{*}$
varies from the critical probabilites
$p^{*}_{c}(\alpha, d)
=1 - \exp{ [-1/(C_{d}(\alpha) 2^{\alpha}) ] }$ to
$p^{*}=1$.
However, if one introduces a more convenient variable,
namely,

\begin{equation}
r^{*} \equiv 1 - \exp{ \left( - \frac{J
\phi(\alpha)}{k_{B} T} \right) }
= 1- (1-p^{*})^{ \frac{\phi(\alpha)}{
N^{*}(\alpha)} }
\end{equation}

\noindent then all these MF probability
percolation curves  for 
different values of $\alpha$ and $d$ coalesce into
a single
curve described by the
equation

\begin{equation}
P_{\infty}(r^{*}) = 1 - (1 -
r^{*})^{P_{\infty}(r^{*})}
\end{equation}

The critical value $r^{*}_{c}$
where $P_{\infty}(r^{*})$ vanishes
is $r_{c}^{*}=1-\exp{(-1)} =
0.63212\ldots$,  which leads to the
MF critical
probability

\begin{equation}
p_{c}(\alpha) = 1 - \exp{\left(
\frac{-1}{\phi(\alpha)} \right)
}
\label{pcdealfa}
\end{equation}

Combining Eqs. (\ref{pcdealfa}),
(\ref{phi02}) and (\ref{n*}) one
verifies that $p_{c}(\alpha \to d^{+})$
vanishes asymptotically
as:

\begin{equation}
p_{c}(\alpha \to d^{+}) \sim
\frac{1}{C_{d}(d) 2^{d}}
 (\frac{\alpha}{d} -1)  
\end{equation}

\noindent
which, in the particular case of $d=1$,
gives

\begin{equation}
p_{c}(\alpha\to 1^{+}, d=1) \sim \frac{1}{2}
(\alpha -1)
\label{pcritico}
\end{equation}

Notice that the asymptotic
behavior (Eq. (\ref{pcritico}))
coincides with the lower bound for
$p_{c}(1<\alpha\le
2,d=1)$\cite{Schulman}.

---------------  
\subsection{Monte Carlo results}

We performed a Monte
Carlo simulation using the heat bath  algorithm on the
one-dimensional
Hamiltonian (\ref{H2}) with $h=0$ and periodic boundary
conditions for
$N=300$, $600$ and $1200$, for $q=2$, $3$ and $5$, and
different values of
$0\leq\alpha< 1$. We calculated the magnetization
per spin (\ref{mag}) as a
function of $T^{*}=T/N^*$ for different system sizes and performed a
numerical
extrapolation
for $1/N \to 0$.

In Fig. 1 we compare the numerical results of the $q=2$ case for $m(T^*)$
vs. $2 k_{B} T^*/ 2^{\alpha} J$ for different values of $\alpha$ with the mean field solution (equation
(\ref{magq2})). We see that all the numerical curves fall into a single one
in excelent agreement with the MF prediction.
\begin{figure}
\centerline{\epsfig{file=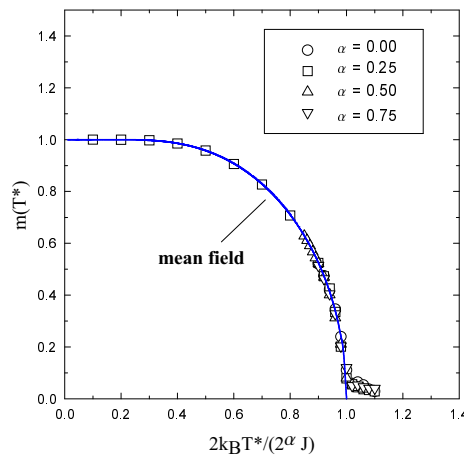,width=6cm}}
\caption{Monte Carlo extrapolated results (symbols) of $m(T^*)$ vs.
$2k_{B}T^*/(2^{\alpha}J)$ compared with the mean field solution
(solid line) for $q=2$}
\label{fig1}
\end{figure}

In Fig. 2  we make the
same comparison for the $q=3$ and $q=5$ cases.
The solid lines represent
the MF solution given by equation (\ref{mfequation}) for $h=0$. The dotted
lines
in this figure indicate the mean field prediction for the critical
temperature jumps (\ref{Tc}). Again we observe, in both cases, an
excelent agreement  between our simulations and MF results for $0 \leq
\alpha < 1$, including the first order phase transition for $q\geq 3$.

\begin{figure}
\centerline{\epsfig{file=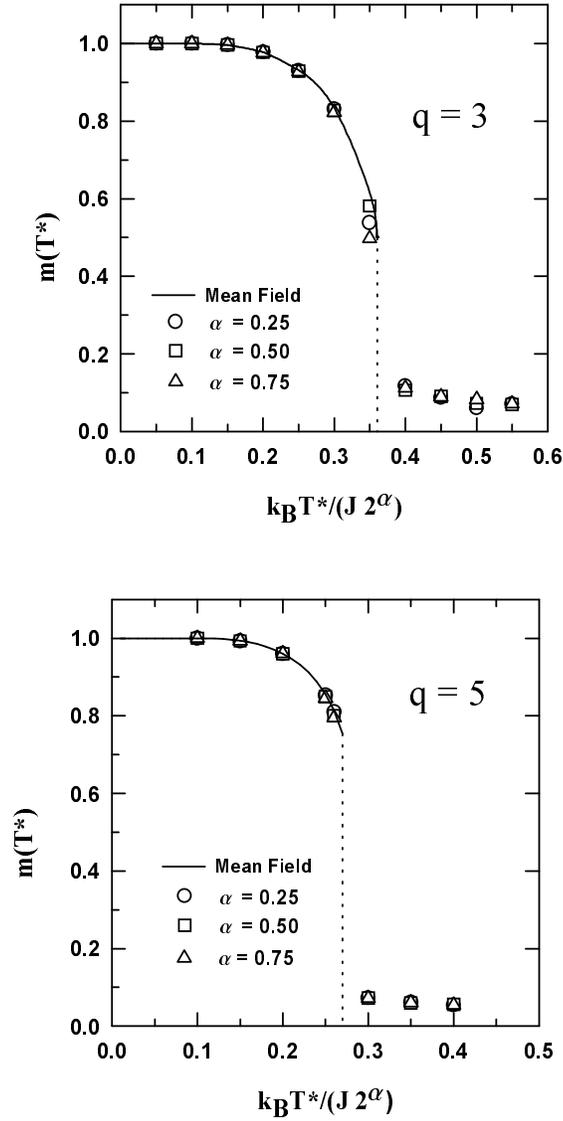,width=14cm}}
\caption{Monte Carlo extrapolated results (symbols) of $m(T^*)$ vs. $
k_{B}T^*/(2^{\alpha}J)$ compared with the mean field solution
(solid line) for $q=3$ and $q=5$.}
\label{fig2}
\end{figure}


\section{Antiferromagnetic Ising model with long-range interactions}

We now consider an Ising model with competing LR antiferromagnetic
and short-range ferromagnetic interactions in an external field, which is
described by the Hamiltonian:

\begin{equation}
H = -J_F \; \sum_{<i,j>}  S_i S_j + J
\;
\sum_{(i,j)}  \frac{1}{r_{ij}^\alpha} \; S_i S_j - h\; \sum_i
S_i
\,\,\,\,\,\,\,\,\, \text{($S_i=\pm 1$\,\, $\forall$
$i$)},
\label{H4}
\end{equation}

\noindent where $J>0$, $J_F>0$ and the sum $\sum_{<i,j>}$ runs over
nearest-neighbor sites of a d-dimensional hypercubic lattice. The above
Hamiltonian reduces, for $J_{F}=1$ and $J=0.5$, to the model studied by 
Sampaio et al \cite{Sampaio} through Monte Carlo simulations. 

A mean field version of this model can be
obtained by considering the
Hamiltonian

\begin{equation}
H_{MF} = - \sum_i
h_{eff}^i \; S_i
\label{HMF2}
\end{equation}

\noindent
with

\begin{equation}
h_{eff}^i = - J\; \sum_{j\neq
i}\frac{1}{r_{ij}^\alpha}\; m_j +
              J_F\; \sum_{j\, nn\, i} m_j
+ h,
\label{heff}
\end{equation}

\noindent where the sum $\sum_{j\, nn\,
i}$ runs over all nearest-neighbor
sites of $i$ and

\begin{equation}
m_j
\equiv \frac{1}{Z_{MF}}\; Tr_{\{ S_i \}} \left\{ 
S_j\; e^{-\beta\,
H_{MF}(\{ S_i
\})} \right\} \; ,
\label{mj}
\end{equation}

\noindent
with

\begin{equation}
Z_{MF} = Tr_{\{ S_i \}}  e^{-\beta\, H_{MF}(\{ S_i
\})}.
\end{equation}

Now we consider the particular case of a square
lattice. Dividing 
our lattice into two square interpenetrated sublattices
A and B we
can propose a solution of the form:

\begin{equation}
m_i =
\left\{\begin{array}{ll}
   m^A& if\;\;\;\;\; i\in A \\
   m^B&
if\;\;\;\;\; i\in B
   \end{array} \right.
.
\label{mi}
\end{equation}

\noindent Let us introduce the
functions

\begin{eqnarray}
\phi^{(1)}(\alpha)& \equiv & \left. \sum_{j\in
A} \frac{1}{r_{ij}^\alpha}
                     \right|_{i\in A} \\
\nonumber
\phi^{(2)}(\alpha)& \equiv & \left. \sum_{j\in B}
\frac{1}{r_{ij}^\alpha}
                     \right|_{i\in A}
\label{phi1}
\end{eqnarray}

\noindent with

\[
\phi^{(1)}(\alpha)+\phi^{(2)}(\alpha)=\phi(\alpha) \]

\noindent It can be
easily seen that

\begin{eqnarray}
\phi^{(1)}(\alpha)&=& 2^{-\alpha/2}\;
\phi(\alpha) \\ \nonumber
\phi^{(2)}(\alpha)&=& (1-2^{-\alpha/2})\;
\phi(\alpha)
\label{phi2}.
\end{eqnarray}

\noindent Then, substituting Eqs.(\ref{mi}) and (\ref{phi1}) into
Eq.(\ref{heff}) we obtain

\begin{equation}
h_{eff}^i = \left\{\begin{array}{ll}
    -J\,
\phi^{(1)}(\alpha)\, m^A - J\, \phi^{(2)}(\alpha)\, m^B +

4\, J_F\, m^B + h & \;\; if \;\;\; i\in A \\
    -J\, \phi^{(2)}(\alpha)\,
m^A - J\, \phi^{(1)}(\alpha)\, m^B +
                      4\, J_F\, m^A +
h & \;\; if\;\;\; i\in B
   \end{array}
\right.
\label{heff2}
\end{equation}

\noindent Combining
Eqs.(\ref{heff2}), Eq.(\ref{HMF2}) and Eq.(\ref{mj}) we arrive, after
some algebra, to the following set of MF equations for the
magnetization $m=m^A+m^B$ and for the staggered magnetization
$m_s=m^A-m^B$:

\begin{equation}
m =  \frac{ sinh\left( 2\beta\, \left[
h-(J\,\phi(\alpha)-4\,J_F)m
       \right]\right)}
   {cosh\left( 2\beta\,
\left[ h-(J\,\phi(\alpha)-4\, J_F)m\right]\right)
  + cosh\left( 2\beta\,
\left[ (J\,\phi(\alpha)\,(2^{1-\alpha/2}-1)  +
4\, J_F)m_s\right]\right)}
\label{m-ms0}
\end{equation}

\begin{equation}
m_s  =  \frac{-sinh\left(
2\beta\, \left[ (J\,\phi(\alpha)\,
           (2^{1-\alpha/2}-1)+4\,
J_F)m_s \right]\right)}
   {cosh\left( 2\beta\, \left[
h-(J\,\phi(\alpha)-4\, J_F)m\right]\right)
  + cosh\left( 2\beta\, \left[
(J\,\phi(\alpha)\,(2^{1-\alpha/2}-1)  +
4\,
J_F)m_s\right]\right)}.
\label{m-ms}
\end{equation}

For $h \neq 0$ it is easy to verify that the only solution of Eq.(\ref{m-ms}) and Eq.(\ref{m-ms0}) is $m_s=0$ and

\begin{equation}
m = tanh\left[ \beta\,
\left(
h-(J\,\phi(\alpha)-4\,J_F)m\right)\right]
\label{meanfieldaf}
\end{equation}

In figure 3 we compare a numerical solution of Eq.(\ref{meanfieldaf}) with
the Monte Carlo data of Sampaio et al\cite{Sampaio} for $\alpha=1$,
$J_F=1$, $J=1/2$, $\beta\, N^*(\alpha)=(0.3)^{-1}$ and different values of
$N$. The function $C_d(\alpha)$ for $d=2$ was evaluated numerically. We obtained, for $\alpha=1$, $C_2(1)=0.8813
\pm 0.0001$. The comparison for other values of $\alpha< 2$ gave similar
results.
 
\begin{figure}
\centerline{\epsfig{file=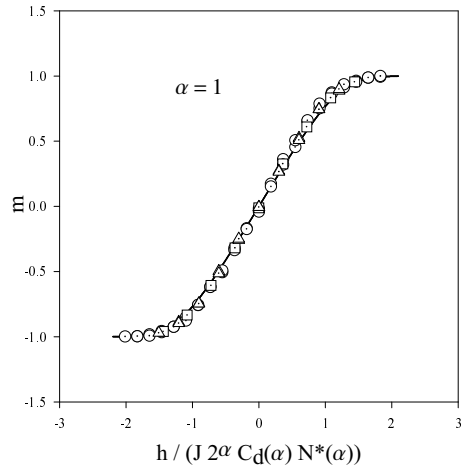,width=6cm}}
\caption{Monte Carlo simulations of Sampaio {\em et al.} 
 for the magnetization $m$ vs. a rescaled magnetic field for square lattice  
 sizes of $32 \times 32$ (circles),  $48 \times 48$ (squares)
 and $64 \times 64$ (triangles) for $\alpha=1$, $J_{F}=1$,
$J=1/2$ and $T^{*}=T/N^{*}=0.3$. The solid line represents the
MF solution given by Eq.(55) for N=64x64.}
\label{fig3}
\end{figure}

A similar comparison for $\alpha=2$ is made in figure 4 ($C_2(2)=0.746\pm0.001$). The Monte Carlo data of
Sampaio et al\cite{Sampaio} do not agree very well with
the MF magnetization, suggesting that corrections to the mean field result
should be taken into consideration in this borderline case (where
$\alpha=d$ ). Notice that  theoretical \cite{Larkin} predictions and
experimental results \cite{Nielsen} obtained for the critical behavior of
$d=3$ uniaxial ferromagnets with exchange and strong dipolar interactions
show that corrections to the MF behavior are needed in this $\alpha=d=3$
case. 

Summarizing this section, we verify that the Monte Carlo
simulations for the
equation of state of the $d=2$ LR Ising antiferromagnet
are in excellent agreement with the MF prediction when $0\leq\alpha<2$.

\begin{figure}
\centerline{\epsfig{file=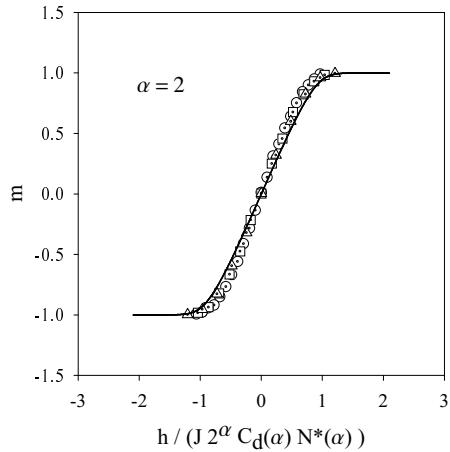,width=6cm}}
\caption{Monte Carlo simulations of Sampaio {\em et al.} 
 for the magnetization $m$ vs. a rescaled magnetic field for square lattice 
 sizes of $32 \times 32$ (circles),  $48 \times 48$ (squares)
 and $64 \times 64$ (triangles) for $\alpha=2$, $J_{F}=1$,
$J=1/2$ and $T^{*}=T/N^{*}=0.3$. The solid line represents the
MF solution given by Eq.(55) for N=64x64.}
\label{fig4}
\end{figure}

\section{Conclusions}
    
We have analyzed, in this paper, two long range spin models with power law 
decaying interactions ($r^{-\alpha}$) under an uniform magnetic field $h$: (i) the $q$-state LR Potts ferromagnet on d-dimensional hypercubic lattices (including the $q\to 1$ and $h\to 0^{+}$ case of LR bond percolation) and (ii) the LR square Ising antiferromagnet with first-neighbor ferromagnetic
interactions. Both models present non-extensive thermodynamic behaviors when
$0 \leq \alpha \leq d$, but their thermodynamic functions become finite
when conveniently scaled variables are  used \cite{Tsallis}. We have derived
this scaling for $q\geq2$ and have shown that the mean field probability
percolation (i.e. the  percolation order parameter) satisfies a similar
scaling. The derived MF solution for the free energy of the LR Potts model
led to spontaneous magnetization curves which agree very well with our
Monte Carlo simulations for $d=1$ and $q=2$, $3$ and $5$ states for
different values of $0\leq\alpha<1$. An excellent agreement occurred
also between the derived MF equation of state for the above $d=2$
antiferromagnetic LR model and the Monte Carlo simulations of Sampaio et al
\cite{Sampaio} for distinct values of $0\leq\alpha<2$. Our results strongly
suggest that the \textbf{mean field theory is exact for LR non-extensive
spin models with} \textbf{$0\leq\alpha<d$}.
This conjecture is also supported by the fact that Bergersen et al  \cite{Bergersen} Monte Carlo simulations of the correlation function of the LR d=1 Ising  ferromagnet reproduce the MF result when $0<\alpha<1$. Since our conjecture predicts, for the $d=1$ LR Potts ferromagnet, a first order transition for $q\geq3$ and $0\leq\alpha<1$, it matches nicely with previous results \cite{Glumac2} exhibiting first order transition for
$1<\alpha<\alpha_{c}(q)$.

Our results show that mean field behaviour is robust against variations of the range of interactions $\alpha$ within the non-extensive region, for a large class of magnetic systems. If our conjecture were true, this would have important practical implications: if you are considering systems with slow enough decaying interactions  then you do not need sophsiticated approximations.

It would be very interesting to extend the present analyis to more general non-extensive systems of interacting particles with long-range interactions.

\section{Acknowledgments}

Fruitful discussions
with Z. Glumac, C. Tsallis and E.M.F.Curado are acknowledged.
We are
greatful to Sampaio and colaborators for sending us their
Monte Carlo data.
This work was partially supported by grants
from Consejo Nacional de
Investigaciones Cient\'\i ficas y T\'ecnicas
CONICET (Argentina), Consejo
Provincial de
Investigaciones Cient\'\i ficas y Tecnol\'ogicas
(C\'ordoba,
Argentina) and Secretar\'\i a de Ciencia y
Tecnolog\'\i a de la
Universidad Nacional de C\'ordoba
(Argentina), Conselho Nacional de
Desenvolvimento Cient\'{\i}fico e
Tecnol\'ogico (CNPq) and
PRONEX/FINEP/MCT.

\end{document}